\documentclass[showpacs,preprintnumbers,amsmath,amssymb]{revtex4}
\usepackage[dvips]{graphicx}
\usepackage{color}
 
\def\setR{\mathbb{R}}

\def\ie {i.e.} 

\newcommand{\norm}[1]{\parallel\!#1\!\parallel}

\newcommand{\dron}[3]{\frac{\partial^{#1} {#2}}{\partial{#3}^{#1}}}

\newcommand{\sss}[1]{\scriptscriptstyle #1}

\begin{document}

\title{Conformal use of retarded Green's functions for the Maxwell field in de~Sitter space}

\author{
S. Faci$^1$, E. Huguet$^1$, J. Renaud$^2$
}
\affiliation{$1$ - Universit\'e Paris Diderot-Paris 7, APC-Astroparticule et Cosmologie (UMR-CNRS 7164), 
Batiment Condorcet, 10 rue Alice Domon et L\'eonie Duquet, F-75205 Paris Cedex 13, France.  \\
$2$ - Universit\'e Paris-Est, APC-Astroparticule et Cosmologie (UMR-CNRS 7164), 
Batiment Condorcet, 10 rue Alice Domon et L\'eonie Duquet, F-75205 Paris Cedex 13, France.
} 
\email{faci@apc.univ-paris7.fr, huguet@apc.univ-paris7.fr, jacques.renaud@univ-mlv.fr}

\date{\today}


\begin{abstract}
We propose a new propagation formula for the Maxwell field in de Sitter space which exploit the conformal invariance of this field together with a conformal gauge condition. This formula 
allows to determine the classical electromagnetic field in the de Sitter space from given 
currents and initial data. It only uses the Green's function of the massless Minkowskian scalar field.
This leads to drastic simplifications in practical calculations. We apply this formula to the classical problem of the two charges of opposite signs at rest at the North and South Poles of the de~Sitter space.
\end{abstract}

\maketitle

\section{Introduction}

In this paper we introduce a propagation formula which
allows to determine the classical electromagnetic field in de Sitter space from given 
currents and initial data. 
We show that this expression takes a simple form 
thanks to a conformal treatment inspired by \cite{pconf3}. 
Indeed, this propagation formula only involves the retarded Green's function 
of the Minkowskian conformal scalar field. One of our objectives is to 
show that this expression simplifies the practical calculations. 

The validity of Green's functions in covariant gauges in de Sitter space
has been matter of debate \cite{Woodard,HiguchiCheong-1} (and references herein). 
In stimulating papers \cite{HiguchiCheong-1,HiguchiCheong-2},  Higuchi, Cheong and Nicholas 
study specific examples. 
In particular, they show that the retarded Green's function obtained from the two-point
function of Allen and Jacobson \cite{AllenJacobson} correctly reproduces the field of two 
opposite charges which are stationary at the North and South Poles of the de Sitter space.

In the present work we show that our propagation formula, which only involves a scalar Green's function, 
leads also to the correct answer. This result has to be compared with that of Higuchi and Cheong \cite{HiguchiCheong-1} in the sense that our method comes also from the quantization of the 
Maxwell field but in an SO$_0(2,4)$-invariant way which uses the Eastwood-Singer \cite{EastwoodSinger} conformal gauge condition instead of the usual Lorenz
condition. But maybe more importantly, that result is obtained by solving the two charges problem 
almost without calculation. This
is a consequence of the formalism introduced for the study of conformally invariant fields on 
de Sitter space \cite{pconf3,pconf1,pconf2}. Initially this formalism has been devised for
the SO$_0(2,4)$-invariant quantization of the free Maxwell field, a quantization we obtained 
in \cite{pconf3}. In fact, this formalism applies to the classical field as well. It leads to a usable propagation formula by specializing in  systems of coordinates in which the Maxwell equations and the gauge condition take their Minkowskian form.

Since the main formula resulting from our study is rather easily implemented and, on the contrary, its
proof needs more explanations we organize the paper broadly into two parts. The first one, Sec. \ref{SecMethExample}, can be viewed as a ``toolbox" for practical calculations. No proofs are given 
there and we propose to the reader to be confident with our claims. We first summarize how the
de~Sitter space are realized from $\setR^6$. In particular we introduce what we call the 
Minkowskian charts, which are basically coordinates systems on the de~Sitter space in which 
conformally invariant expressions take their Minkowskian form. Then, we gives an overview of 
the classical Cauchy problem for the Maxwell field in this context.
That section ends with the 
application to the aforementioned problem of the two opposite charges located at the North and 
South Poles. The second part, Sec.  \ref{SecFormalism}, is divided in two main subparts. 
In the first place we remind the geometrical framework 
and complement it. In a second place
we adapt the method used in our previous work \cite{pconf3} to a classical situation where currents are present, and finally,  give the proof of the propagation formula. 
There, we show in particular that only the Minkowskian massless scalar Green's function is needed 
in the propagation formula. 
Some explicit expressions are given for reference in appendix \ref{AppExplicit}. 
A new proof, which differs from that of \cite{HiguchiCheong-1}, for a general 
propagation formula (not specialized in de Sitter) is given in \ref{AppPropGen}.

\subsection*{Conventions and notations}
Here are the conventions for indices:
\begin{eqnarray*}
\alpha, \beta, \gamma, \delta, \ldots &=&0, \ldots, 5,\\
\mu,\nu,\rho,\sigma\ldots &=&  0, \ldots, 3,\\
i, j, k, l, \ldots &=& 1, \ldots, 3,\\
I,J, \ldots&=& c, 0,\ldots,3,+.
\end{eqnarray*} 
The coefficients of the metric $\mbox{diag}(+,-,-,-,-,+)$ of $\setR^{6}$ are denoted $\tilde{\eta}_{\alpha \beta}$.
For convenience we set $\eta_{\mu\nu} := \tilde{\eta}_{\mu\nu}$. 

Various spaces and maps are used throughout this paper. Except otherwise stated, quantities related to $\setR^6$ and its null cone $\mathcal{C}$ are labeled with a tilde, those defined on the
de~Sitter space are denoted with a super or subscript $H$.

We will uses four Minkowskian charts to cover the de~Sitter space
(see Sec. \ref{SecRevGeom} and \ref{SecGeometry}) they are called North, South, Down North and
Down South. The quantities related to these charts appears with a super or subscript $M$ which
can take the values $\{N, S, DN, DS\}$. These systems are thus denoted $\{x_{\sss M}\}$, we set 
$r_{\sss M} = \vert \boldsymbol{x}_{\sss M}\vert$.
In each of these charts the de Sitter space is related to a Minkowski space through a Weyl rescalling:
$g^{\sss H}_{\mu\nu}(x_{\sss M})\nolinebreak=\nolinebreak K_{\sss M}^2(x_{\sss M}) \, \eta_{\mu\nu}$ (\ref{SecMinkCoordRev}).
We shall call Minkowskian counterpart of some desitterian field $F^{\sss H}$ the 
field $F^{\sss M} = K_{\sss M}^{-s} F^{\sss H}$, where $s$ is the so-called conformal 
weight of the field F.

\section{The calculation method}\label{SecMethExample}

\subsection{A conformal realization of the de~Sitter and Minkowski manifolds}\label{SecRealisdSHonCs}
Roughly speaking, two ideas underly our geometrical construction: the former is that working on 
the five dimensional null cone $\mathcal{C}$ of $\setR^6$, which is invariant 
under the conformal group SO$_0(2,4)$, allows to preserve the symmetries rather easily 
(in particular the invariance under the de~Sitter subgroup), the later is to exploit the fact that
there is a relation between points located on different sections of the cone but belonging to
the same half-lines issued from the origin of $\setR^6$. 
\subsubsection{Brief review of the geometrical construction}\label{SecRevGeom}
\begin{figure}
\begin{center}
\includegraphics[width = 7.5cm]{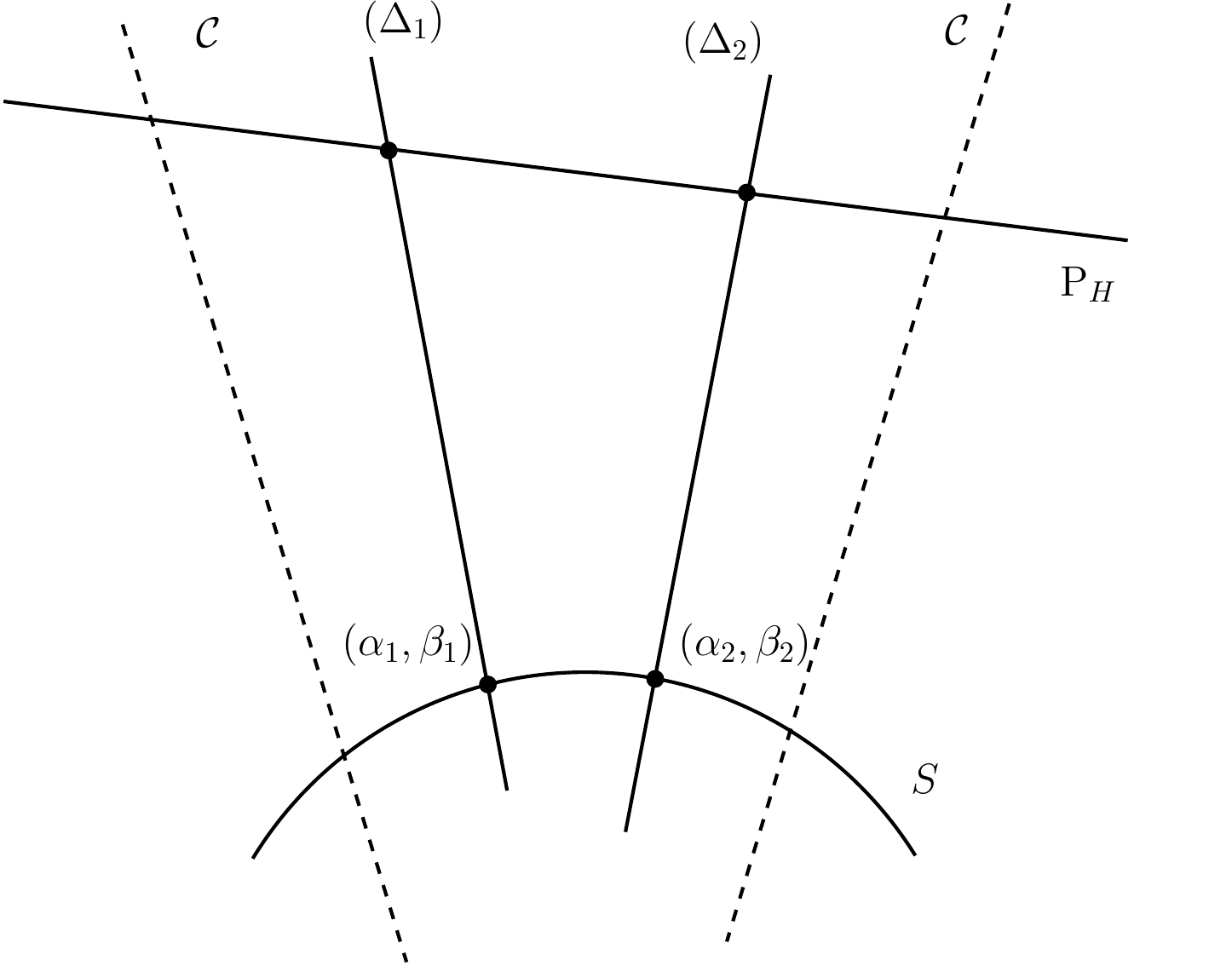}
\caption{The main construction: To each point of $X_{\sss H} = P_{\sss H}\cap {\cal C}$ corresponds a unique half-line $\Delta$, that is a point of ${\cal C'}$ which can be located, in coordinates 
$(\alpha, \beta)$, on the intersection $\mathcal{C_{\sss S}}$  of ${\cal C}$ with the sphere $S$.}
\label{Fig1}
\end{center}
\end{figure}

The construction we describe is pictured in Fig. \ref{Fig1}.
The de~Sitter space is obtained as the intersection $X_{\sss H}$ of the five 
dimensional null cone $\mathcal{C}$ of $\setR^6$ 
and a moving plane $P_{\sss H}$.  The positive parameter $H$ being related to the Ricci  
scalar $R$ through $R = 12 H^2$.  Since each point of the intersection $X_{\sss H}$ is intercepted by a 
half-line of the cone (a generatrix line), the manifold $X_{\sss H}$ can be realized  as a subset of the set $\mathcal{C'}$ of the half-lines of the cone. 
Now, every half-line intercepts also the euclidean unit $5$-sphere $S$ of $\setR^6$ once. This allows us to 
realize the abstract manifold $\mathcal{C'}$ as the intersection $\mathcal{C_{\sss S}}$ of the unit $5$-sphere $S$ and the cone $\mathcal{C}$, each half-line of $\mathcal{C}$ corresponding to a point of $\mathcal{C_{\sss S}}$. It follows that the de~Sitter space is realized as projections $X^{\sss S}_{\sss H}$ along the half-lines of $\mathcal{C}$ onto $S$ of the spaces $X_{\sss H}$.
In the sequel for convenience and readability we shall call de~Sitter space both the realization $X_{\sss H}$ and the central projection $X_{\sss H}^{\sss S}$ onto $\mathcal{C_{\sss S}}$. Note that,
from the point of view of group theory the manifold $X_{\sss H}$ is invariant under the de Sitter group SO$_0(1,4)$ suppr. Note also that the linearity of the action 
of SO$_0(2,4)$ on $\setR^6$ ensures that there is a natural action of this group on ${\mathcal C}'$ (see \cite{pconf1} 
for details).

\begin{figure}
\begin{center}
\includegraphics[width = 7.5cm]{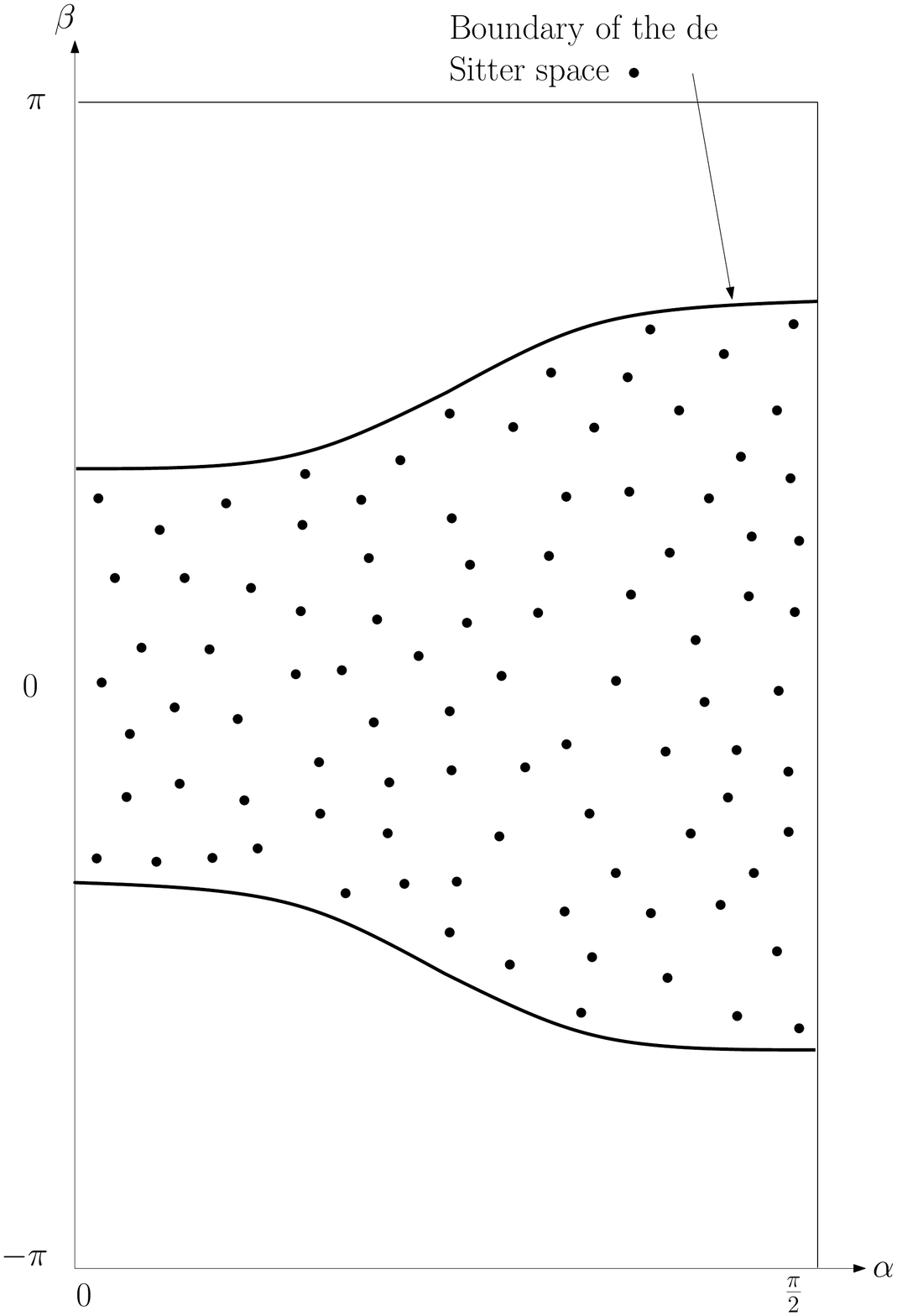}
\caption{
Realization of the abstract manifold $\mathcal{C'}$: the intersection $\mathcal{C_{\sss S}}$ of the cone $\mathcal{C}$ and the $5-$sphere $S$ is the whole rectangle. The projection of the de~Sitter space on $S$, for a given value of $H$,  is the dotted region. The North and South Poles are respectively the part of the lines $\alpha = 0$ and $\alpha = \pi$ which bound the dotted region.
}\label{Fig2}
\end{center}
\end{figure}

The calculation of the line element on $X_{\sss H}$ shows a local Weyl relation between the de~Sitter space $X_{\sss H}$ and 
$\mathcal{C_{\sss S}}$:
\begin{equation}\label{WeylXHCS}
ds_{\sss H}^2 = \Omega_{\sss H}^2 ds_{\sss S}^2, 
\end{equation}
where $ds_{\sss H}^2$ and $ds_{\sss S}^2$ are respectively the line elements of $X_{\sss H}$ and $S$ inherited from the metric of $\setR^6$. This allows us, in particular, to obtain the very convenient 
conformal diagram of Fig. \ref{Fig2} in which the projection $X^{\sss S}_{\sss H}$ can be represented. 
The boundaries of $X^{\sss S}_{\sss H}$ are the set of points of $\mathcal{C_{\sss S}}$ such that $\Omega_{\sss H} \rightarrow \infty$. This diagram is built using the $\alpha,\beta$ system of 
coordinates on $\mathcal{C_{\sss S}}$:
\begin{equation}\label{coordAlpBetsurC}
\left \{
 \begin{array}{lcl}
  y^5 &=&  \cos \beta \\
  y^0 &=&  \sin \beta \\
  y^i &=&  \sin \alpha\; \omega^{i}(\theta,\varphi)\\
  y^4 &=&  \cos \alpha,
 \end{array}
\right .
\end{equation}
in which $\beta\in[-\pi,\pi[,\, \alpha,\theta\in[0,\pi], \varphi\in[0,2\pi]$ 
and $\omega^{i}(\theta, \varphi)$ correspond to the usual spherical coordinates on $S^2$. In this coordinate system, the Weyl factor $\Omega_{\sss H}$ reads
\begin{equation}
\Omega_{\sss H}=  \frac{ 1 }{ (1-H^2)\cos{\alpha} + (1+H^2)\cos{\beta} }.
\end{equation}

\subsubsection{Minkowskian coordinates}\label{SecMinkCoordRev}

In what follows, we use four different charts of $\mathcal{C_{\sss S}}$ 
which are generically called Minkowskian coordinate systems. In these systems
the de~Sitter and the Minkowski metric are related through the Weyl rescaling:
\begin{equation}\label{WeylSystMink}
g^{\sss H}_{\mu\nu} = K_{\sss M}^2 \eta_{\mu\nu},
\end{equation}
where $K_{\sss M}$ is the Weyl factor related to the chart. 

\begin{figure}
\begin{center}
\includegraphics[width = 8.cm]{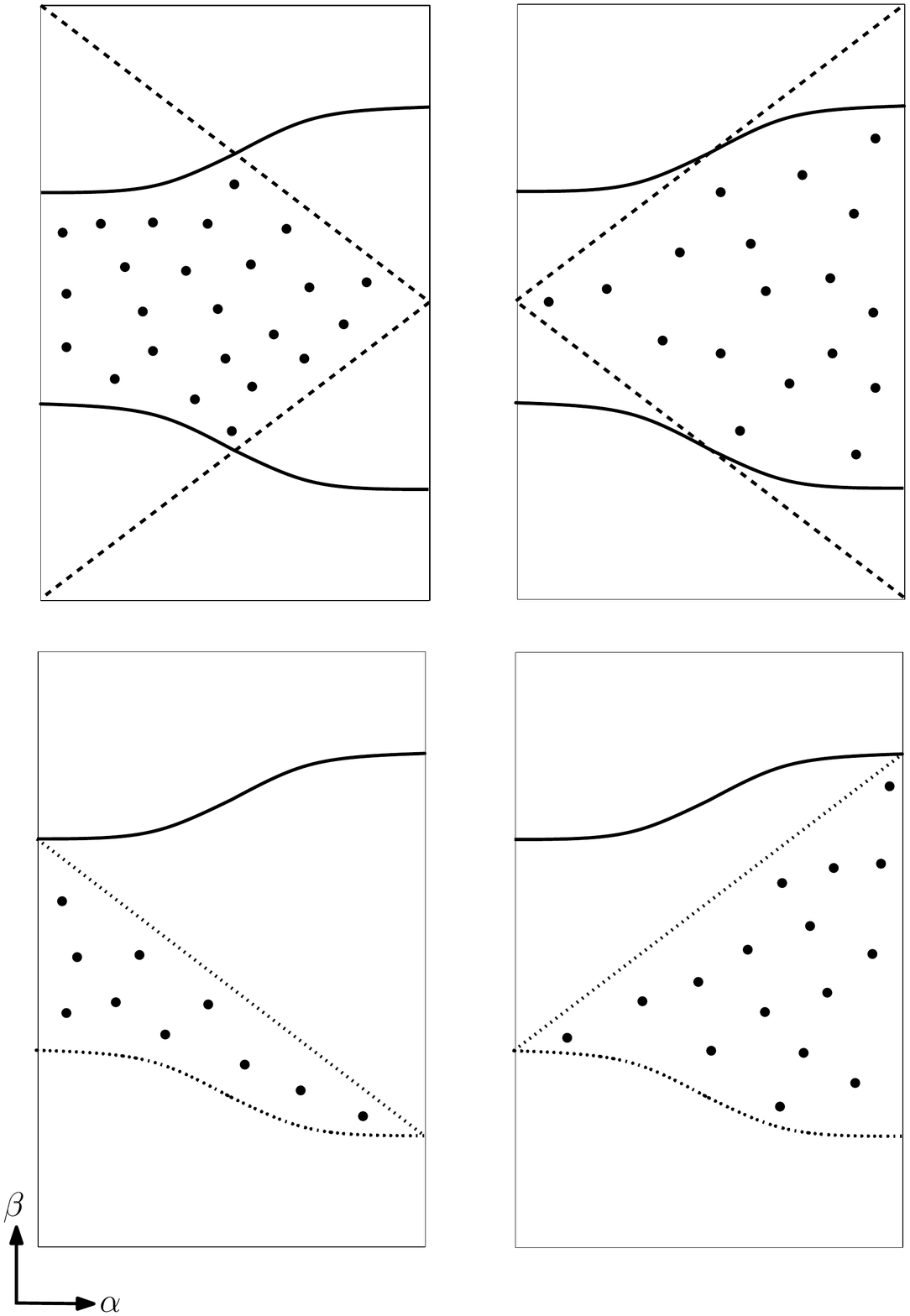}
\caption{The four Minkowskian charts used in this paper: up left, the North chart with the coordinates $\{x_{\sss N}\}$, up right, the South chart with the coordinates $\{x_{\sss S}\}$, 
down left, the Down North chart with the coordinates $\{x_{\sss DN}\}$ and  down right, the Down South chart with the coordinates $\{x_{\sss DS}\}$. The dots indicate the region 
covered by the chart. These regions depend on the value of $H$. Note that, by contrast with the North and South charts, the whole past infinity of $X_{\sss H}$ 
is always a boundary of the $\{x_{\sss DN}\}$ and $\{x_{\sss DS}\}$ charts. }\label{Fig3}
\end{center}
\end{figure}

They are the North chart $\{x_{\sss N}\}$, the South chart $\{x_{\sss S}\}$, the Down North $\{x_{\sss DN}\}$ and Down South $\{x_{\sss DS}\}$ charts. They are introduced in Sec. \ref{SecMinkChart-R6}, additional 
formulas are collected in appendix \ref{AppSystem}.
The regions covered by each of these charts are shown in Fig. \ref{Fig3}. 
The two charts $\{x_{\sss N}\}$ and $\{x_{\sss S}\}$ together cover the whole de~Sitter space.

For our study, in which conformal invariance plays an important role, these 
Minkowskian charts dramatically simplify the practical calculations: 
in such a chart an equation which is
invariant under the Weyl rescaling (\ref{WeylSystMink}) takes its Minkowskian form.

\subsection{Propagation of the electromagnetic field}\label{SectPropEMFiel}

In this section we give the method to calculate the classical Maxwell field in 
presence of currents in de~Sitter space within the conformal approach built in a 
previous paper \cite{pconf3} for the free quantum field. The adaptation of that
formalism to the present situation is detailed in Sec.  \ref{SecFormalism}. Here
we focus on the practical application of the results which are stated without 
proof. These are given in Sec. \ref{SecFormalism}.

The method can be summarized as follows. We assume that initial data are given on some 
Cauchy surface and that the currents in de~Sitter space are known. We choose an atlas 
composed of Minkowskian charts. In each chart, in addition to the Maxwell equations, 
we choose a gauge condition which is invariant under the Weyl rescaling given in 
(\ref{WeylSystMink}). We then apply, as detailed in Sec. \ref{SectProOneChart} below, 
the propagation formula which is very simple in this context. 
In each chart we thus obtain a field $A^{\sss H}$. In regions in which the charts overlap the
fields determined in each chart only differ  by a pure gauge term. Thus, the Faraday tensor $F_{\mu\nu}$ 
is the same on the whole de~Sitter space.

\subsubsection{Propagation in a Minkowskian chart}\label{SectProOneChart}

In de~Sitter space we consider the Maxwell equation and the gauge condition
\begin{equation}\label{MaxAndDMA}
\left \{
\begin{aligned}  
&\square_{\sss H} A_\mu^{\sss H} - \nabla_\mu \nabla A^{\sss H} + 3H^2 A_\mu^{\sss H} = J_\mu^{\sss H}\\
&\nabla^\mu\left(\nabla_\mu + W_\mu^{\sss M} \right)\left(\nabla^\nu - 
 W^{{\sss M}\nu}\right)A_\nu^{\sss H} = 0,
\end{aligned}
\right . 
\end{equation} 
where $W^{\sss M}$ is the one-form $W^{\sss M} = \mathrm{d} \ln K_{\sss M}^2$ of components 
$W^{\sss M}_\mu = \nabla_\mu \ln K_{\sss M}^2$. 
This one form is thus related to the  Minkowskian chart in which the local Weyl rescaling between
de~Sitter and Minkowski space is given by (\ref{WeylSystMink}). Remark that, although the charts considered in Sec. \ref{SecMinkCoordRev} are singular at some points, it can be checked that $W^{\sss M}$ is regular, i.e. the second line of the formula (\ref{MaxAndDMA}) is well defined (and covariant) on the whole de Sitter space.
Nevertheless, for practical calculation this expression is usable only on the corresponding chart in which it takes a simple form.

In the Minkowskian chart considered the above system reduces to
\begin{equation}\label{MaxAndDNAMink}
\left \{
\begin{aligned}
&\partial^2 A^{\sss H}_\mu - \partial_\mu \partial  A^{\sss H}  = K_{\sss M}^2  J_{\mu}^{\sss H} \\
&\partial^2 \, \partial  A^{\sss H}  = 0,
\end{aligned}
\right .
\end{equation}
where we have set
$\partial^2 := \eta^{\mu\nu}\partial_\mu \partial_\nu$, 
$\partial  A^{\sss H} = \eta^{\mu\nu}\partial_\mu A_\nu^{\sss H}$ 
although we are still in de Sitter space.

Now, the system (\ref{MaxAndDNAMink}) can also be obtained through the  Weyl rescaling (\ref{WeylSystMink}) on the system (\ref{MaxAndDMA}). Indeed, using the fact that the conformal weight of the Maxwell form field $A^{\sss H}$ under a Weyl rescaling is zero, a direct calculation shows that the gauge condition is invariant under (\ref{WeylSystMink}). For the Maxwell equation this invariance is satisfied if the conformal weight of the current $J_{\mu}^{\sss H}$ is equal to $2$. We thus define the Minkowskian counterpart $J^{\sss M}_{\mu}$ of $J_{\mu}^{\sss H}$ in such a way that the invariance under  (\ref{WeylSystMink}) be satisfied, namely 
\begin{equation}\label{lesJ}
J^{\sss M}_{\mu} :=  K_{\sss M}^2  J_{\mu}^{\sss H}.
\end{equation}
Note that $J^{\sss M}_{\mu}$ does not  correspond, in general,  to a physical current 
in the Minkowski space. 

Using the above properties, the propagation formula for $A^{\sss H}$ in the Minkowskian
chart considered reads
\begin{equation}\label{source-initial-A-intro}
A^{\sss H}_\mu(x) = A^{(s)}_\mu(x) + A^{(i)}_\mu(x),
\end{equation}
with a ``source" term related to the current
\begin{equation*}
A^{(s)}_\mu(x) =  \int_{D^+(\Sigma)}\!\!\!\!\!\! dx'^4  \  G^{\sss R}_{s}(x,x') \ J_{\mu}^{\sss M}(x'),
\end{equation*}
where ${D^+(\Sigma)}$ is the causal future of $\Sigma$,
and an ``initial" term related to the initial data
\begin{equation*}
A^{(i)}_\mu(x) =   \int_{\Sigma} \sigma^{\mu'} \  G^{\sss R}_{s}(x,x') 
\stackrel\leftrightarrow \partial_{\mu'} 
A^{\sss H}_{\mu}(x'),
\end{equation*}
where 
$G^{\sss R}_{s}(x,x')$ is the usual Minkowskian massless scalar retarded function (see for instance
\cite{BarutED}):
\begin{equation}\label{GreenScalMinkIntro}
G^{\sss R}_{s}(x,x')=\frac{\delta(x^0-x'^0 - \norm{\boldsymbol{x} - \boldsymbol{x'}})}{4 \pi 
\norm{\boldsymbol{x} - \boldsymbol{x'}}} \ \theta(x^0 - x'^0) ,
\end{equation}
and $ \sigma^{\mu'} $ is the normal element of surface of the Cauchy surface $\Sigma$ 
considered as a sub-manifold of the underlying Minkowskian space.
This calculation is valid as soon as the intersection of the past cone of $x'$ with the future of $\Sigma$ is included in the Minkoskian chart considered. We analyze this point in connection with the propagation
in the whole de~Sitter space below.

\begin{figure}
\begin{center}
\includegraphics[width = 7.5cm]{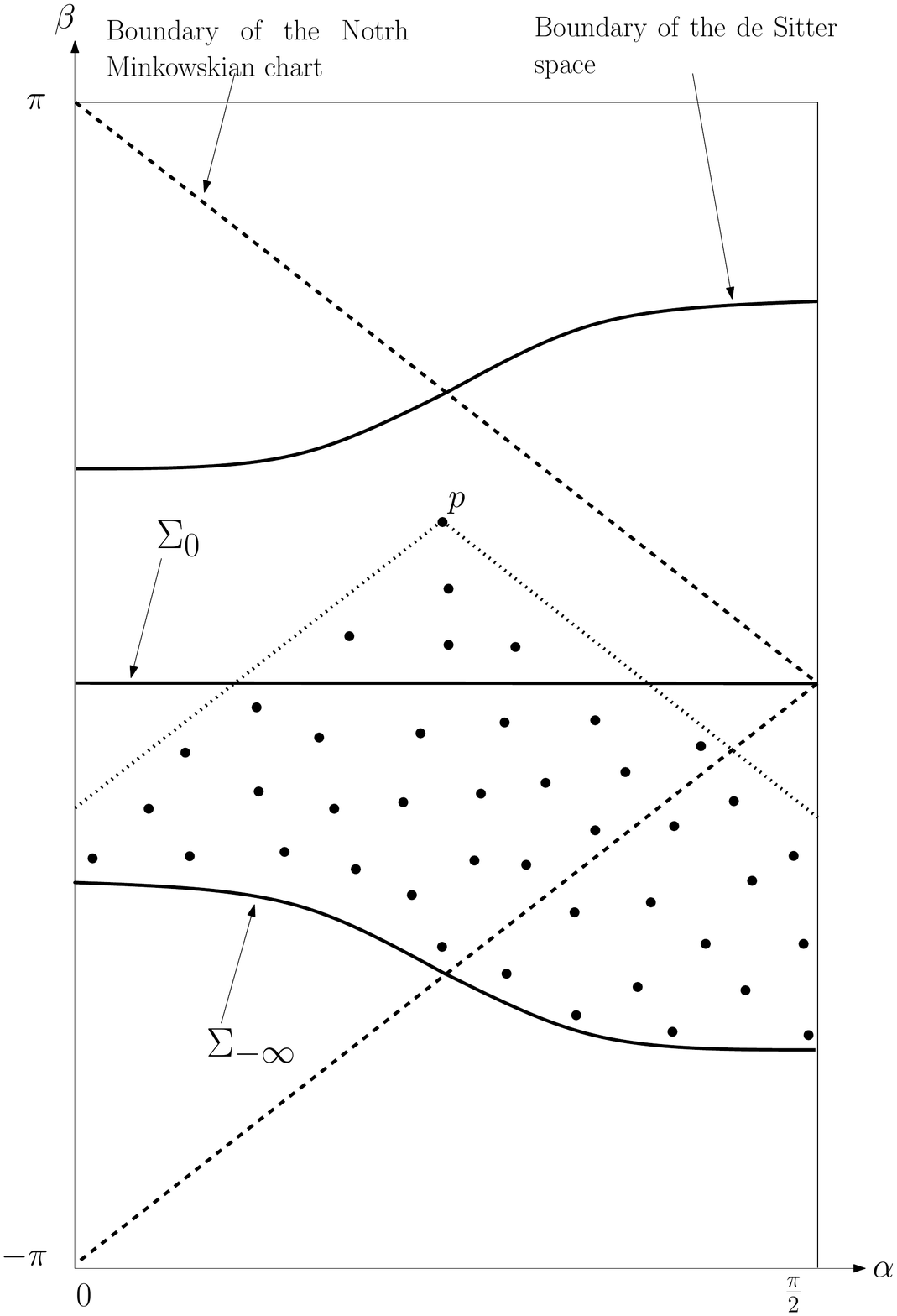}
\caption{
Use of the propagation formula in connection with the Minkowskian charts. On this example the field 
$A^{\sss H}$ at the point $p$ can be calculated using the propagation formula in the North Minkowskian chart $\{x_{\sss N}\}$ if the data are given on $\Sigma_0$. 
Indeed the intersection of the causal past of $p$ (doted region) with the future of $\Sigma_0$ is included into the North chart. This is no more the case if the initial data are given for instance on the Cauchy surface at the past infinity $\Sigma_{-\infty}$. In this case another chart must be used.
}\label{Fig4}
\end{center}
\end{figure}

\subsubsection{Propagation in the whole de~Sitter space}\label{PropIndS}

The previous paragraph gives a propagation formula in one Minkowskian chart, if one consider a problem on the whole de~Sitter manifold, at least two charts are required. In that case, one has to proceed by steps, using the propagation formula several times.

We will consider two examples illustrating the method. 
They can be visualized with the help of the Fig. (\ref{Fig4}). In the first example 
we assume that the initial data are given on the
Cauchy surface $\Sigma_0$ defined through $\beta=0$ (that is to say $x^0_{\sss N}=x^0_{\sss S}=0$). The knowledge of the field $A_\mu^{\sss H}$ and its derivatives on $\Sigma_0$ and the source on the future of $\Sigma_0$  allows to compute the field everywhere in the future of $\Sigma_0$, using the formula (\ref{source-initial-A-intro}) in the North chart or in the South chart depending on position of the point $p$ Fig. (\ref{Fig4}) and (\ref{Fig3}).
Even more, using the advanced Green function, one can compute the field everywhere in the past of $\Sigma_0$.

The second example is the following. Suppose that one knows the source and the values of the field on the past infinity: $\Sigma_{-\infty}$ and  the source currents are known in the future of $\Sigma_{-\infty}$. The Cauchy surface $\Sigma_{-\infty}$ is defined in both Down North and Down South charts through $x^0_{\sss DN}=x^0_{\sss DS}=0$.
Then, one can obtain the field on the whole Cauchy surface $\Sigma_0$ using these two charts. After 
what, one can obtain the field everywhere using the calculation from $\Sigma_0$ described above.

We must note that the usability of the method for practical calculations require that the initial data are given on a Cauchy surface contained in some Minkowskian charts that allow the above ``step by step"  calculation.

\subsection{A specific example: two opposite charges in de~Sitter}

The simplest configuration we can study in de Sitter space is that of two charges of opposite signs stationary at opposite Poles. This configuration is reminiscent of the problem of one stationary charge in the Minkowski space with the difference that in a spatially compact space, as the de~Sitter space, the total charge must vanish \cite{BrillDeser}. Another known feature related to the topology of the de~Sitter space is concerned by the Cauchy problem \cite{Penrose}: an initial field must be specified on the past infinity $\Sigma_{-\infty}$, even if this surface is outside the causal future of both charges, 
this is mainly a consequence of the Gauss theorem. The Cauchy problem related to this distribution is comprehensively reviewed in the article of Bi\u{c}\'{a}k and Krtous   \cite{Bicak-Krtous-1}. The explicit calculation using the retarded Green's function built from the de~Sitter invariant two-point function obtained by Allen and Jacobson \cite{AllenJacobson} 
(in Lorenz gauge) is considered by  Higuchi, Cheong and Nicholas \cite{HiguchiCheong-1,HiguchiCheong-2} and by Woodard \cite{Woodard}. 

Here, we show that under the same initial conditions we can reproduce the electromagnetic field compatible with this two charges distribution in the whole de Sitter space using the propagation formula (\ref{source-initial-A-intro}). The benefit of the method proposed here is that the calculations are reduced to those of the Coulombian field (also considered in \cite{HiguchiCheong-1}) in Minkowski space, \ie~for one charge at rest.

To be precise, in the conformal global coordinate system ($\tau, \chi$) 
 used in \cite{HiguchiCheong-1}, and linked to the ($\alpha, \beta$) system through
\begin{equation}\label{global-north}
\left \{
 \begin{aligned}
 \frac{\sin \tau}{ \cos \tau} &=  \frac{ H\sin \beta }{ (1-H^2)\cos{\alpha} + (1+H^2)\cos{\beta} } \\
\frac{\sin \chi}{\cos \tau} &=   \frac{ H \sin \alpha }{ (1-H^2)\cos{\alpha} + (1+H^2)\cos{\beta} }.
  \end{aligned}
\right.
\end{equation}  
the non-zero component of the field generated by the two opposite charges $\pm q$ stationary at the 
North Pole ($\chi=0$) and South Pole 
($\chi=\pi$) reads

\begin{equation}\label{solution-globale}
A_{\tau}  = \frac{q}{4\pi} \ \cot\chi.
\end{equation}
This solution, despite its slightly different form, is the solution used by 
Woodard \cite{Woodard} and also, up to a gauge term, that of Higuchi and Cheong \cite{HiguchiCheong-1}. 
Note that, under the transformation $\chi \rightarrow \pi - \chi$, which interchanges the two poles, 
the field simply changes its sign which shows that the solution (\ref{solution-globale}) is indeed 
that of two charges of opposite sign.

\subsubsection{Propagating from $\Sigma_{-\infty}$}\label{SecPropz}

We thus consider the same Cauchy problem as Higuchi and Cheong \cite{HiguchiCheong-1}, in which 
initial data are given at the past null infinity. 

In order to propagate the field from $\Sigma_{-\infty}$, we have to consider the charts Down North 
$\{x_{\sss DN}\}$ and Down South $\{x_{\sss DS}\}$ and write the field (\ref{solution-globale}) in these coordinate systems. Then, up to a gauge term the solution to be reproduced (\ref{solution-globale}) reads:
\begin{equation}\label{solution-nord-lambda}
A^{\sss H}_{\mu} (x_{\sss DN}) =  \delta_{\mu}^0 \ \frac{q}{4\pi r_{\sss DN}},
\end{equation}
in the Down North chart and
\begin{equation}\label{solution-sud-lambda}
A^{\sss H}_{\mu} (x_{\sss DS}) =  \delta_{\mu}^0 \ \frac{-q}{4\pi r_{\sss DS}},
\end{equation}
in the Down South chart with similar notations.

In Down North chart the current reads
\begin{equation*}
J_{\mu}^{\sss H} (x_{\sss DN}) = +q \ \delta_{\mu}^{\sss 0} K^{-2}_{\sss DN} \delta^3 (\boldsymbol{x}_{\sss DN}),
\end{equation*}
where $K_{\sss DN}$ is the Weyl factor of Eq. (\ref{WeylXHCS}) for the Down North chart,
and hence, thanks to (\ref{lesJ})
\begin{equation*}
J_{\mu}^{\sss M} (x_{\sss DN}) = +q \ \delta_{\mu}^{\sss 0}  \delta^3 (\boldsymbol{x}_{\sss DN}).
\end{equation*}
The Cauchy's surface is parametrized by $x_{\sss DN}=0$.
The initial incoming field on ($\Sigma_{-\infty}$) is given by 
\begin{equation*}
A^{\sss H}_{\mu} (x_{\sss DN})\big\vert_{\sss \Sigma_{-\infty}} =  \delta_{\mu}^0 \ \frac{q}{4\pi r_{\sss DN}}, \quad \partial A^{\sss H}(x_{\sss DN})\big\vert_{\sss \Sigma_{-\infty}}=0.
\end{equation*}

Now, inserting the above quantities in the propagation formula (\ref{source-initial-A-intro}), one 
obtains
\begin{equation*}
\begin{aligned}
A^{(s)}_\mu (x_{\sss DN}) = & \int_{D^+({x^0}'_{\sss DN}=0)}\!\!\!\!\!\!dx'^4_{\sss DN}  \ J_{\mu}^{\sss 0}(x'_{\sss DN}) \  G^{\sss R}_{s}(x_{\sss DN},x'_{\sss DN}) \\
= & \frac{q}{4\pi r_{\sss DN}} \ \delta_{\mu}^{\sss 0} \ \theta(x^0_{\sss DN} - r_{\sss DN}),
\end{aligned}
\end{equation*}
and
\begin{equation*}
 \begin{aligned}
A_{\mu}^{(i)}(x_{\sss DN}) = & \int_{x'^0_{\sss DN}=0}\!\!\!\!\!\! dx'^3_{\sss DN}  \  G^{\sss R}_{s}(x_{\sss DN},x'_{\sss DN})\  \stackrel\leftrightarrow{\partial'}_{0}\  A_{\mu}^{\sss H}(x'_{\sss DN})\\
= &  \frac{q}{4\pi r_{\sss DN}} \ \delta_{\mu}^{\sss 0} \ \theta(r_{\sss DN}-x^0_{\sss DN}).
\end{aligned}
\end{equation*}
Note that, in view of the expressions of $G^{\sss R}$ and of $J_{\mu}^{\sss M}$, the first line of these expressions are formally identical to that obtained for a single charge in Minkowski space. This 
avoids the calculation since the result is already known in that case.
Summing the source and the initial parts, we find that the generated electromagnetic field in the Down 
North chart takes the expected form (\ref{solution-nord-lambda}). 

The calculation in the Down South chart is identical up to the sign of the charge, 
and the replacement of the index $DN$ by $DS$.
Calculating and summing the source and the initial parts in the South chart, we find that the generated electromagnetic field takes the expected value (\ref{solution-sud-lambda}).

The propagation formula works as expected in the part of the space covered by 
the Down North and
Down South charts. To obtain the field on the whole de~Sitter space, one has to proceed by step 
as in the second example of Sec. \ref{SectPropEMFiel}. To this end we can use Cauchy surface $\Sigma_0$ on which the field is now known This is done in the next section.

\subsubsection{Propagating from $\Sigma_{0}$}

In order to obtain the initial field on $\Sigma_0$, we have just to express the field (\ref{solution-sud-lambda}) and (\ref{solution-nord-lambda}) in the North or South chart.

In the North Chart, the above solutions (\ref{solution-sud-lambda}) and (\ref{solution-nord-lambda}) 
reads up to a gauge term
\begin{equation}\label{coulomb-field-N}
A_{\mu}^{\sss H} (x_{\sss N}) =  \frac{+q}{4\pi r_{\sss N}} \ \delta_{\mu}^{\sss 0}.
\end{equation}

In the South Chart, we obtain
\begin{equation}\label{coulomb-field-S}
A_{\mu}^{\sss H} (x_{\sss S}) =  \frac{-q}{4\pi r_{\sss S}} \ \delta_{\mu}^{\sss 0}.
\end{equation}

Consider now a point $p(x_{\sss N})$ in the future of $\Sigma_0$ and in the North chart, see Fig. (\ref{Fig4}). 
The current is the North chart reads 
\begin{equation}\label{source-current-N}
J_{\mu}^{\sss H} (x_{\sss N}) = +q \ \delta_{\mu}^{\sss 0} K^{-2}_{\sss N} \delta^3 (x_{\sss N}).
\end{equation}

Using the propagation formula (\ref{source-initial-A-intro}), the source part reads
\begin{equation}\label{source-N}
 \begin{aligned}
A^{(s)}_\mu(x_{\sss N}) = & \int_{D^+(x'^0_{\sss N}=0)}\!\!\!\!\!\! dx'^4_{\sss N}  \ J_{\mu}^{\sss M}(x'_{\sss N}) \  G^{\sss R}_{s}(x_{\sss N},x'_{\sss N}) \\
= & \frac{+q}{4\pi r_{\sss N}} \ \delta_{\mu}^{\sss 0} \ \theta(x^0_{\sss N} - \frac{2}{H} - r_{\sss N}).
\end{aligned}
\end{equation}

The initial part reads
\begin{equation}\label{initial-N}
 \begin{aligned}
A^{(i)}_\mu(x_{\sss N}) = & \int_{x'^0_{\sss N}=0}\!\!\!\!\!\! dx'^3_{\sss N}  \ 
G^{\sss R}_{s}(x_{\sss N},x'_{\sss N}) 
\  \stackrel\leftrightarrow{\partial'}_{0}\ A_{\mu}(x'_{\sss N}) \\
= & \frac{+q}{4\pi r_{\sss N}} \ \delta_{\mu}^{\sss 0} \ \theta(r_{\sss N} - x^0_{\sss N} + \frac{2}{H}).
\end{aligned}
\end{equation}
Once again the calculations are exactly the same as the Minkowskian ones,
and the generated electromagnetic field in the North chart reads
\begin{equation*}
A^{\sss H}_\mu(x_{\sss N}) =  \frac{+q}{4\pi r_{\sss N}} \ \delta_{\mu}^{\sss 0},
\end{equation*}
as expected (\ref{coulomb-field-N}).

Following the same steps for a point $p(x_{\sss S})$ in the future of $\Sigma_0$ and in the South chart
we can conclude that the electromagnetic field is well reproduced.

\subsubsection{Comments}

The propagation formula works as expected in the whole space, even though we need several charts to map the whole de Sitter space. This drawback is balanced by the simplicity of the calculations.
One can note that these are very similar to that appearing in the
Minkowskian problem at the beginning of the work of Higushi and Cheong \cite{HiguchiCheong-1}.
This analogy is due to the complete exploitation of the conformal invariance of 
Maxwell field including the gauge condition (\ref{MaxAndDMA}).

\section{Justification of the calculation method}\label{SecFormalism}
This section is devoted to the formalism which justifies the calculation method, and in particular
the propagation formula, used in the previous section. We first set some geometrical tools, then 
we describe how the equations (\ref{MaxAndDMA}) for the potential $A^{\sss H}$ are related
to a set of constrained scalar equations which can be solved using a scalar Green's function. 

\subsection{The geometrical framework}\label{SecGeometry}

\subsubsection{Realization of the de~Sitter space}

Our geometrical construction has been summarized in Sec. \ref{SecRealisdSHonCs}~: the de~Sitter space is first realized as the intersection $X_{\sss H}$ of the five dimensional null cone $\mathcal{C}$ and a hyper-plane $P_{\sss H}$ defined through $f_{\sss H}(y) = 1$ where
\begin{equation}\label{fH-def-y}
f_{\sss H}(y)=\frac{1}{2}\left((1 + H^2) y^{5} + (1 - H^2) y^{4}\right).
\end{equation}
Then, $X_{\sss H}$ is projected along the half-lines issued from the origin of $\setR^6$ onto  
the intersection $\mathcal{C_{\sss S}}$ of the unit $5$-sphere $S$ and the cone $\mathcal{C}$. This
intersection $\mathcal{C_{\sss S}}$ is a realization of the abstract set of the half-lines 
$\mathcal{C'} $, the cone modulo the dilations. The de~Sitter space is thus considered as a subset 
of $\mathcal{C'} $ realized on $\mathcal{C_{\sss S}}$.

\subsubsection{The Minkowskian charts}\label{SecMinkChart-R6}
The construction of the previous paragraph can be used to realize other spaces than the de~Sitter space on the same underlying set $\mathcal{C_{\sss S}}$. As a consequence, common coordinate systems can be used to locate points on different spaces in the regions where they overlap.  In particular the Minkowskian hyperplanes $P_{\sss M}$, defined through $f_{\sss M}(y) = 1$  with  
\begin{align}
f_{\sss N}(y) &=+ \frac{1}{2}\left(y^{5} + y^{4}\right),\label{fN-def-y}\\
f_{\sss S}(y) &= +\frac{1}{2} H^2 (y^{5} - y^{4}),\label{fS-def-y}\\
f_{\sss DN}(y) &= +\frac{1}{2}\left( y^{5} + y^{4} - H^2 (y^{5} - y^{4})\right) - H y^0,\label{fDN-def-y}\\
f_{\sss DS}(y) &= -\frac{1}{2}\left( y^{5} + y^{4} - H^2 (y^{5} - y^{4})\right) - H y^0,\label{fDS-def-y}
\end{align}
are spaces whose central projections onto $\mathcal{C_{\sss S}}$ are the subsets corresponding 
respectively to the North, South, Down North and Down South charts appearing in Fig. \ref{Fig3}. 
Each of these spaces can be endowed with a Minkowskian and Cartesian coordinate system. Each of these systems is the restriction to the realization of the space as a subset of $\mathcal{C_{\sss S}}$ of two systems defined in $\setR^6$. The need for two systems in $\setR^6$ for one Minkowskian system will become clear
when we will define the auxiliary fields in Sec.\ref{SecDetildage}. The explicit relations defining these systems of coordinates in $\setR^6$ are 
\begin{equation}\label{coordMinkGene-xHxN}
\left \{
 \begin{array}{lcl}
  x^c_{\sss M} &=&  {\displaystyle \frac{y^\alpha y_\alpha}{4 f_{\sss M}^2}}\\
  x^0_{\sss M} &=&  {\displaystyle \zeta_{\sss M}(y) \frac{y^0}{f_{\sss M}}} \\
  x^i_{\sss M} &=&  {\displaystyle \frac{y^i}{f_{\sss M}}}\\
  x^+_{\sss H} &=& f_{\sss H},~~\mathrm{or}~~x^+_{\sss M} = f_{\sss M}
 \end{array}
\right .
\end{equation}
with 
\begin{align}
\zeta_{\sss N}(y) =& \,\zeta_{\sss S}(y) = 1, \label{zeta-N-S}\\
\zeta_{\sss DN}(y) =& \,\zeta_{\sss DS}(y) =\frac{f_{\sss H}}{H y^0}.
\end{align}
The Minkowskian coordinates are obtained for $x^c_{\sss M} = 0$, which is the restriction to the cone $\mathcal{C}$, and $x^+_{\sss H} = 1$ (respectively $x^+_{\sss M} = 1$) which is the restriction 
to the plane $P_{\sss H}$ (respectively $P_{\sss M}$).
That is, $x^c_{\sss M} = 0$ and $x^+_{\sss H} = 1$ gives the Minkowskian charts on the 
de~Sitter space, and $x^c_{\sss M} = 0$ and $x^+_{\sss M} = 1$ gives the Minkowskian chart 
on the corresponding Minkowskian space.
Some practical formulas regarding these coordinates are given in appendix \ref{AppSystem}.

\subsubsection{Homogeneous functions on the cone}

Let us now remind a property which is of central importance for our construction. 

Let $f_{\sss \mathcal{V}}$ be a function defined on $\setR^6$ homogeneous of degree one. Let $\mathcal{V}$ be the intersection of the null cone $\mathcal{C}$ and the hyper-surface defined 
through $f_{\sss \mathcal{V}}(p) = 1$, $p\in\setR^6$. Let $p$ be a point of the cone,
$p^{\sss \mathcal{C'}}$ the 
half-line linking $p$ to the origin of $\setR^6$ and 
$p^{\sss \mathcal{V}}$ the point located at the intersection of the hyper-surface with the half-line. 
Then 
\begin{equation*}
p^{\sss \mathcal{V}} = p/f_{\sss \mathcal{V}}(p).
\end{equation*}
This is easily proved since
\begin{equation*}
f_{\sss \mathcal{V}}\left(\frac{p}{f_{\sss \mathcal{V}}(p)}\right) 
= \frac{1}{f_{\sss \mathcal{V}}(p)} f_{\sss \mathcal{V}}(p) = 1.
\end{equation*}
Then, for any function $\widetilde{F}$ homogeneous of degree $r$, on the cone $ \mathcal{C}$, we 
define a function $F^{\sss \mathcal{V}}$ on the cone modulo the dilations $\mathcal{C'}$ such that
\begin{equation}\label{tildeGeneral}
F^{\sss \mathcal{V}}(p^{\sss \mathcal{C'}}):=\widetilde{F}(p^{\sss \mathcal{V}}) 
= \left(\frac{1}{f_{\sss \mathcal{V}}(p)}\right)^r\widetilde{F}(p).
\end{equation}
This definition means that the value of the function $F^{\sss \mathcal{V}}$ at  the
half-line containing a point $p^{\sss \mathcal{V}}$ of the manifold $\mathcal{V}$ is that of 
$\widetilde{F}$ at this point.

Applying the above relation for $f_{\sss \mathcal{V}} = f_{\sss H}$ and 
$f_{\sss \mathcal{V}} = f_{\sss M}$ 
(\ref{fH-def-y})-(\ref{fDS-def-y}) one obtains 
\begin{equation}\label{FH-FM}
F^{\sss H}(p^{\sss \mathcal{C'}}) = \left(K_{\sss M}(p^{\sss \mathcal{C'}})\right)^r
\, F^{\sss M} (p^{\sss \mathcal{C'}}), 
\end{equation}
where we defined
\begin{equation}\label{K}
K_{\sss M}(p^{\sss \mathcal{C'}}) = \frac{f_{\sss M}(p)}{f_{\sss H}(p)},\; p\in p^{\sss \mathcal{C'}}.
\end{equation}
Note that since the two function $f_{\sss M}$ and $f_{\sss H}$ are
homogeneous of the same degree their ratio only depends on the position of
the half-line which contains the point $p$, not on the position 
of the point $p$ itself. 

Since $p^{\sss \mathcal{C'}}$ can be located on $\mathcal{C_{\sss S}}$ the 
relation (\ref{tildeGeneral})  can be translated in all charts of the de 
Sitter space (realized on $\mathcal{C_{\sss S}}$) 
introduced in Sec. \ref{SecMinkChart-R6}. Direct calculations in these systems shows that 
$K_{\sss M}$ is nothing but the Weyl factor between the metrics of the de Sitter space and
the Minkowskian chart $M$, the degree of homogeneity $r$ becoming the conformal weight.

\subsection{Maxwell equations and conformal gauge in presence of current}\label{SecMaxwellEqs}
In this section we adapt to our present problem the formalism used in our previous work \cite{pconf3}.
There are two differences: the field is classical and currents are present. On the one hand this 
situation is more simple because the auxiliary fields needed to achieve a covariant quantization in
\cite{pconf3} are now only classical constraints which mainly fix the gauge condition and allows to
obtain Maxwell field on de~Sitter space from a field defined in $\setR^6$. On the other hand, the 
introduction of currents gives to the problem a global character which forces us to use different gauges
in different charts on de~Sitter space if one wants to take benefit of the simplifications induced by the 
SO$_0(2,4)$-invariant construction from $\setR^6$, in particular a propagation 
formula (\ref{source-initial-A-intro}) which uses a scalar Green's function.

\subsubsection{The fields from $\setR^6$}\label{SecDetildage}

The scheme starts from the Dirac's six cone formalism  \cite{Dirac6cone,MS} in which   
we consider the SO$_0(2,4)$-invariant inhomogeneous wave equation 
\begin{equation}\label{eqa}
\Box_{6} \, \tilde a_\alpha =\tilde j_\alpha,
\end{equation}
where $\Box_6 := \tilde{\eta}^{\gamma\beta}\tilde{\partial}_\gamma\tilde{\partial}_\beta$ and 
$\tilde a = \tilde a_\alpha dy^\alpha$ and  $\tilde j = \tilde j_\alpha dy^\alpha$ are both one-forms 
in $\setR^6$. Following \cite{Dirac6cone,MS}, the one-forms $\tilde a$ and $\tilde{j}$ must be 
homogeneous of degree $-1$ and $-3$ respectively in order to consider the field and the equation 
on the cone $\cal C$. 

Since both $\tilde a$ and $\tilde{j}$ are homogeneous, the formula (\ref{tildeGeneral}) allow us to 
define $a^{\sss H}$ and $j^{\sss H}$ on the de~Sitter space.
The equation (\ref{eqa}) can then be pushed on de~Sitter space where it reads 
\begin{equation}\label{Box+H2aH}
\left(\square_{\sss H}^s + 2 H^2\right)a^{\sss H}_\alpha= j^{\sss H}_\alpha,
\end{equation}
where $\square_{\sss H}^s$ is the desitterian Laplace-Beltrami operator acting on a 
scalar, this means that in this equation each component  $a^{\sss H}_\alpha$ must be considered 
as a scalar.

The same considerations apply to define through (\ref{tildeGeneral}) the Minkowskian fields 
$a^{\sss M}$ and $j^{\sss M}$ from $\tilde a$ and $\tilde{j}$ and to write the equation (\ref{eqa}) 
for these fields on a Minkowskian chart. 
One obtains
\begin{equation}\label{BoxM-aM}
\partial_{\sss M}^2 a^{\sss M}_\alpha= j^{\sss M}_\alpha,
\end{equation}
where 
${\displaystyle \partial_{\sss M}^2 = \eta^{\mu\nu} \dron{}{~}{x^\mu_{\sss M}}\dron{}{~}{x^\nu_{\sss M}}}$.
Note that, this amounts to use the Weyl invariance of the scalar equation (\ref{Box+H2aH}). 

Now, let us introduce the sets of auxiliary fields 
$\{\widetilde{A}^{\sss (H)}_c, \widetilde{A}^{\sss (H)}_\mu,
\widetilde{A}^{\sss (H)}_+\}$ and 
$\{\widetilde{A}^{\sss (M)}_c, \widetilde{A}^{\sss (M)}_\mu,
\widetilde{A}^{\sss (M)}_+\}$. These are defined through the
decompositions
\begin{align}
\tilde{a}
&= \widetilde{A}^{\sss (H)}_c dx^c_{\sss M} 
+\widetilde{A}^{\sss (H)}_\mu dx^\mu_{\sss M} + 
\frac{\widetilde{A}^{\sss (H)}_+}{x^+_{\sss H}} dx^+_{\sss H}, \label{decomposition-a-AH}\\ 
\tilde{a}
&= \widetilde{A}^{\sss (M)}_c dx^c_{\sss M} 
+\widetilde{A}^{\sss (M)}_\mu dx^\mu_{\sss M} + 
\frac{\widetilde{A}^{\sss (M)}_+}{x^+_{\sss M}} dx^+_{\sss M}. \label{decomposition-a-AM}
\end{align}
The two decompositions are related to the coordinates (\ref{coordMinkGene-xHxN}). From the first set of auxiliary fields we will obtain the Maxwell field on the de~Sitter space. The second set of auxiliary fields is connected to coordinate systems which, from the considerations of Sec. \ref{SecMinkChart-R6}, corresponds to a projection onto a Minkowskian plane $P_{\sss M}$. This set of fields will give the Maxwell field on a Minkowski space corresponding to the chart $M$. The two sets are related by an extended Weyl transformation defined below.

All the $\widetilde{A}$'s being homogeneous (of degree $0$ in the variable $x^+_{\sss H}$ for the $\widetilde{A}^{\sss (H)}$ and $x^+_{\sss M}$ for the $\widetilde{A}^{\sss (M)}$) 
one can define thanks to (\ref{tildeGeneral}) the auxiliary 
fields on de~Sitter and Minkowski spaces. For convenience we set 
\begin{align}
A^{\sss H}&= A^{\sss (H) H},\label{tildeA-AH}\\
A^{\sss M}&= A^{\sss (M) M}.\label{tildeA-AM}
\end{align}

The above decompositions (\ref{decomposition-a-AH}-\ref{decomposition-a-AM}) induces linear 
relations between the components $\tilde{a}_\alpha$ of $\tilde{a}$ in the $\{dy\}$
basis and the components of the auxiliary fields $\widetilde{A}^{\sss (H)}$ and $\widetilde{A}^{\sss (M)}$
in their respective associated basis. These linear relations can also be translated thanks to (\ref{tildeGeneral}) in linear 
relations between the components of $a^{\sss H}$ and those of $A^{\sss H}$ defined on the de~Sitter space, and the components of $a^{\sss M}$ and those of $A^{\sss M}$ defined on the Minkowskian charts, they are given in appendix \ref{AppRelAa-N}. 

As mentioned above, the $A^{\sss H}$'s are related to each set of $A^{\sss M}$ through a relation 
which we call an extended Weyl transformation. It can be obtained for each pair of auxiliary fields
from the relation
\begin{equation}\label{Weyl-aHaM}
a^{\sss H} = K_{\sss M}^{-1} a^{\sss M},
\end{equation}
which follows from (\ref{FH-FM}) and from the linear relations between the $a$'s and the $A$'s (see 
appendix \ref{AppRelAa-N}). For each set $\{A^{\sss M}\}$ the extended Weyl transformation reads

\begin{equation}\label{WeylExtended}
A^{\sss H}_{\sss I} = A^{\sss M}_{\sss I} +  \left[\frac{\partial~~}{\partial x_{\sss M}^{\sss I}} 
\ln\left(\frac{f_{\sss M}}{f_{\sss H}}\right)\right] A^{\sss M}_+.
\end{equation}  
Note that, $f_{\sss M}/f_{\sss H}$ being homogeneous of degree 0, it does not depend on $x_{\sss M}^+$. 
As a consequence $A^{\sss M}_+ = A^{\sss H}_+$, this implies that the constraint $A_+ = 0$,
which has to be imposed in order to recover the set of Maxwell equations (see \cite{pconf3} for details) 
applies in all charts at the same time.

Finally, the auxiliary set of currents $J^{\sss H}$ and  $J^{\sss M}$ are introduced in the 
same way through the expansion of  $\tilde{j}$ on the two basis 
$\{dx^c_{\sss M},dx^\mu_{\sss M},dx^+_{\sss H}\}$ and $\{dx^c_{\sss M},dx^\mu_{\sss M},dx^+_{\sss M}\}$.

\subsubsection{The Maxwell equations in the North chart}\label{SecEqsNorth}
We now describe how the equations (\ref{MaxAndDMA}) for the potential $A^{\sss H}$ are obtained
from the set of constrained scalar equations (\ref{BoxM-aM}) that can be solved using a scalar Green's function. 
For simplicity we works in the system $\{x^\mu_{\sss N}\}$ (the North chart): it means that $M=N$ in the remainder of Sec. \ref{SecMaxwellEqs}. The generalization
to the other Minkowskian charts will be given in Sec. \ref{SecPropGeneral}.

We follow the same steps as in \cite{pconf3} where the Reader is referred for details.
From Sec. \ref{SecDetildage} the equation in $\setR^6$ (\ref{eqa}) reduces in de Sitter space
to (\ref{Box+H2aH}). Using the relations (\ref{a(A)-H}) between $a^{\sss H}$ and $A^{\sss H}$
leads to a system of differential equations
involving the auxiliary fields. 
In the case of the free fields the Maxwell equations (and a gauge relation) are obtained by
using the additional constraint $y^\alpha\tilde a_\alpha=0$, this comes from the Dirac's formalism. 
Here, this constraint is carried by the auxiliary field $A^{\sss H}_+ = y^\alpha\tilde a_\alpha$ 
which has been introduced originally in the quantum context (\cite{Bayenetal, pconf3} for details). 
When currents are present we are free to impose the additional constraint 
$J^{\sss H}_+ = y^\alpha\tilde j_\alpha =0$. With these two constraints, the equations (\ref{Box+H2aH})
rewrite
\begin{equation}\label{SystAuxA+J+=0-H}
\left \{
\begin{aligned}
&\partial^2 A^{\sss H}_\mu - \partial_\mu \partial  A^{\sss H}  = K_{\sss N}^2  J_{\mu}^{\sss H} \\
&\partial^2 \, \partial  A^{\sss H}  = - K_{\sss N}^2 J_c^{\sss H}\\
& A^{\sss H}_c = - \partial A^{\sss H}.
\end{aligned}
\right .
\end{equation}
The third equation allows to determine $A^{\sss H}_c$ in function of $A^{\sss H}$, the first and the
second equations are just the Maxwell equations and the conformal gauge condition with an additional term
$J_c^{\sss H}$ (compare to (\ref{MaxAndDNAMink})). These two equations can be rewrited 
(see \cite{pconf3}) as (\ref{MaxAndDMA}) with an additional term $J_c^{\sss H}$, precisely
\begin{equation}\label{Maxwell2}
\left \{
\begin{aligned}  
&\square_{\sss H} A_\mu^{\sss H} - \nabla_\mu \nabla A^{\sss H} + 3H^2 A_\mu^{\sss H} = J_\mu^{\sss H}\\
&\nabla^\mu\left(\nabla_\mu + W^{\sss N}_\mu \right)\left(\nabla^\nu - 
 W^{{\sss N}\nu}\right)A_\nu^{\sss H} = - K_{\sss N}^{-2} J_{c}^{\sss H}.
\end{aligned}
\right . 
\end{equation} 
We will prove in the following that we can, without loss of generality, take $J_{c}^{\sss H}=0$. Hence, the solution of the Maxwell equations in the conformal gauge (\ref{MaxAndDMA}) can be obtained 
from the solutions of (\ref{Box+H2aH}).

Now, the same considerations applies to Minkowskian counterparts $A^{\sss N}$ and $a^{\sss N}$. 
That is the solution of Maxwell equations in the conformal gauge in Minkowski space 
\begin{equation}\label{MaxAndDNAMink-N}
\left \{
\begin{aligned}
&\partial^2 A^{\sss N}_\mu - \partial_\mu \partial  A^{\sss N}  = J_{\mu}^{\sss N} \\
&\partial^2 \, \partial  A^{\sss N}  = 0,
\end{aligned}
\right .
\end{equation}
which are, following the consideration of Sec. \ref{SectProOneChart} in Weyl relation with (\ref{MaxAndDMA}),
and can be obtained from the solutions of (\ref{BoxM-aM}).

Finally, reminding that the conformal weight of the Maxwell form field $A^{\sss H}$ under a 
Weyl transformation is zero, that is $A^{\sss H} = A^{\sss M}$, one concludes that 
the solution $A^{\sss H}$ of the Maxwell equations in de~Sitter space in the conformal gauge (\ref{MaxAndDMA}) can be obtained from the solution $a^{\sss N}$ of the equations (\ref{BoxM-aM})
as announced.

\subsubsection{The Propagation formula in the North chart}\label{SecPropFor-N}

In this section we give the proof of the propagation formula (\ref{source-initial-A-intro}). We continue 
to work in the North chart leaving the generalization to Sec. \ref{SecPropGeneral}. In this section, we write $x_{\sss N}=x$ for convenience.

As seen in the previous section the Maxwell equations and the gauge condition are equivalent 
(up to the constraints encoded in the auxiliary field $A^{\sss H}_+=A^{\sss N}_+$ and  
$J^{\sss H}_+=J^{\sss N}_+$) to the
equations (\ref{BoxM-aM}) for the $a^{\sss N}$ repeated here for convenience
in the North chart
\begin{equation}\label{BoxN-aN}
\eta^{\mu\nu} \dron{}{~}{x^\mu}\dron{}{~}{x^\nu} a^{\sss N}_\alpha = j^{\sss N}_\alpha.
\end{equation}
 
Precisely, the field $A^{\sss H} = A^{\sss N}$ can be determined from the $a^{\sss N}$'s 
thanks to the relations 
\begin{equation}\label{solutionA(a)}
A^{\sss N}_\mu(x) =  a_\mu^{\sss N}(x) - 
\frac{1}{2} \eta_{\mu\nu}x^\nu \left(a_5^{\sss N}(x) - a_4^{\sss N}(x) \right),
\end{equation}
which follows from the relation between the $a^{\sss N}$'s and $A^{\sss N}$'s 
(see appendix \ref{AppExplicit}).
Similar relations holds for the current fields
\begin{equation}\label{solutionJ(j)}
J_\mu^{\sss N}(x) =  j_\mu^{\sss N}(x) - 
\frac{1}{2}  \eta_{\mu\nu}x^\nu \left(j_5^{\sss N}(x) - j_4^{\sss N}(x) \right).
\end{equation}

Thus, using the general propagation formula of appendix \ref{AppPropGen} for each component
$a_{\alpha}^{\sss N}$ will give the propagation formula for the field $A^{\sss H}_\mu=A^{\sss N}_\mu$.
In addition,
since each $a^{\sss N}_\alpha$ component behaves, as solution of (\ref{BoxN-aN}), like a 
scalar field,  this is the Green's function for the Minkowskian conformal scalar $G^{\sss R}_{s}$ given in (\ref{GreenScalMinkIntro}) which must be used in the propagation formula.
One may note that, ignoring this last property of the $a_{\alpha}^{\sss N}$'s with respect 
to the equation (\ref{BoxN-aN}) and starting from the Wightman two-point function obtained in 
\cite{pconf3} to derive the retarded Green's function for the $a_{\alpha}^{\sss N}$'s would 
lead to the same result.
Taking into account these considerations, the general formula (\ref{GeneralPropa}) reads
\begin{equation}\label{propagation-a}
a_{\alpha}^{\sss N} = a_{\alpha}^{(s)} + a_{\alpha}^{(i)},
\end{equation}
in which the source part reads
\begin{equation}\label{source-a}
\begin{aligned}
 a_{\alpha}^{(s)} (x)
 & = (j^{\sss N}_{\alpha} * G^{\sss R}_{s}) (x) \\
 & =\int_{D^+(\Sigma)}\!\!\!\!\!\!\!\!\!\!d^4x' \ j^{\sss N}_{\alpha}(x') \ G^{\sss R}_{s}(x,x'),
 \end{aligned}
\end{equation}
and the initial part reads
\begin{equation}\label{initial-a}
\begin{aligned}
 a_{\alpha}^{(i)} (x)
 & = \langle a^{\sss M}_{\alpha}, G^{\sss R}\rangle_s(x)\\
 & =  \int_{\Sigma} \sigma_0^{\nu'}(x') \ G^{\sss R}_{s}(x,x') 
 \stackrel\leftrightarrow \partial_{\nu'} a_{\alpha}^{\sss N}(x'),
  \end{aligned}
\end{equation}
where $\langle ,\rangle_s$ is the Minkowskian Klein-Gordon scalar product and  $G^{\sss R}_{s}$ 
is the Minkowskian conformal scalar retarded Green's function (\ref{GreenScalMinkIntro}) and where $ \sigma^{\nu'}_{\sss 0} $ is the normal element of surface on $\Sigma$ viewed as a sub-manifold of the underlying Minkowskian space.

The next step of the proof consists in expressing the terms $a_{\alpha}^{\sss N}(x')$ and $j^{\sss N}_{\mu}(x')$ in (\ref{source-a}) and (\ref{initial-a}) in terms of $A^{\sss N}(x')$ and $J^{\sss N}(x')$, taking into account the constraints $J_+^{\sss N} = 0$,  $A^{\sss N}_{+}=0$ and $A^{\sss N}_{c}=-\partial A^{\sss N}$. This reads
\begin{equation}\label{solutiona(A)}
\begin{aligned}
&a_5^{\sss N}-a_4^{\sss N}=2A^{\sss N}_c\\
&a_\mu^{\sss N}=\frac{1}{2}\eta_{\mu\nu}x^\nu A^{\sss N}_c+A^{\sss N}_\mu,
\end{aligned}
\end{equation}
and
\begin{equation}\label{solutionj(J)}
\begin{aligned}
&j_5^{\sss N}-j_4^{\sss N}=2J^{\sss N}_c\\
&j_\mu^{\sss N}=\frac{1}{2}\eta_{\mu\nu}x^\nu J^{\sss N}_c+J^{\sss N}_\mu.
\end{aligned}
\end{equation}
The propagation formula for $A^{\sss H}_{\mu}$ then reads
\begin{equation}\label{source-initial-A}
A^{\sss H}_\mu(x) = A^{\sss N}_\mu(x) = A^{(s)}_\mu(x) + A^{(i)}_\mu(x),
\end{equation}
with the source part
\begin{equation}\label{source-A}
\begin{split}
A^{(s)}_\mu(x) &= \int_{D^+(\Sigma)}\!\!\!\!\!\!\!\!\!\!d^4x' 
 G^{\sss R}_s(x,x') \Bigl( J^{\sss N}_\mu(x')  \\
&+\eta_{\mu\nu}\frac{x^{ \nu} - {x'}^{\nu}}{2} \, J_{c}^{\sss N}(x')  \Bigr),
\end{split}
\end{equation}
and the initial part
\begin{equation}\label{initial-A}
\begin{split}
A^{(i)}_\mu(x)  &= 
\int_{\Sigma}\!\!\! \sigma^{\nu'}_{\sss 0} \
G^{\sss R}_s(x,x') \stackrel{\leftrightarrow} {\partial_{\nu'}} 
\Bigl( A^{\sss H}_\mu(x') \\
&-
\eta_{\mu\nu}\frac{x^{ \nu} - {x'}^{\nu}}{2}  \, \partial A^{\sss N}(x')\Bigr).
\end{split}
\end{equation} 

Up to now,  the current $J^{\sss N}_{c}$ remains undefined. Indeed, the term in which it appears in (\ref{source-A}) can be gauged away. To prove this statement, we consider the term
\begin{equation}\label{pure-gauge}
\bar A_\mu(x) = \int_{D^+(\Sigma)}\!\!\!\!\!\!\!\!\! d^4x' 
 G^{\sss R}_s(x,x')\,\eta_{\mu\nu}\frac{x^{ \nu} - {x'}^{\nu}}{2} \, J_{c}^{\sss N}(x') .
\end{equation}
A straightforward calculation, shows that the corresponding Faraday tensor $F_{\mu\nu}=\partial_{\mu} \bar A_{\nu}-\partial_{\nu} \bar A_{\mu}$  is always vanishing. This proves our claim at the end of the previous section: we can, without loss of generality, choose $J_{c}^{\sss H}=K_{\sss N}^2 J_{c}^{\sss N} =0$.

The initial part (\ref{initial-A}) can also be simplified by choosing the additional gauge condition  $\partial A^{\sss N}=0$. 
Note that, this expression reads $(\nabla-W)A^{\sss H}=0$ in a covariant form and together with 
$J_{c}^{\sss H}=0$ is compatible with the conformal gauge condition in (\ref{MaxAndDMA}).

The propagation formula thus simplifies and takes, as claimed, a purely Minkowskian form 
using only the scalar conformal Green's function. The source part gives
\begin{equation}
A^{(s)}_\mu(x) =  \int_{D^+(\Sigma)}\!\!\!\!\!\! dx'^4  \  G^{\sss R}_{s}(x,x') \ J_{\mu}^{\sss N}(x'),
\end{equation}
and the initial part reads
\begin{equation}
A^{(i)}_\mu(x) =   \int_{\Sigma} \sigma^{\mu'} \  G^{\sss R}_{s}(x,x') 
\stackrel\leftrightarrow \partial_{\mu'} 
A_{\mu}(x') .
\end{equation}

Finally, to apply the above formula we need the initial data on some Cauchy surface $\Sigma$ and 
the current $J_\mu^{\sss H}$ in de~Sitter space from which we deduce the Minkowskian counterpart $J_\mu^{\sss N}$.  The point $x$ at which the field is computed must be such that the intersection of the past cone of $x$ with the future of $\Sigma$ is entirely contained in the North chart (Sec. \ref{SecMethExample}).

\subsection{The propagation formula on the whole de~Sitter space}\label{SecPropGeneral}

We will now explain how to obtain the same propagation formula in the other charts, the so-called $S$, $DN$ and $DS$ charts.

In each other chart one would proceed as in the North chart, this would give, in particular,  
relations equivalent to (\ref{solutionA(a)}) between the Maxwell field $A^{\sss H}_\mu$ and 
the corresponding Minkowskian field $a^{\sss M}$. From these relations
one could then obtain a propagation formula following the same steps as before. 
A more convenient and unifying way is to remark that the formulas (\ref{solutionA(a)}), (\ref{solutionJ(j)}), (\ref{solutiona(A)}) and (\ref{solutionj(J)}) are fulfilled with a new set of fields and currents
$\bar a_\alpha^{\sss M}$, $\bar j^{\sss M}$ which are obtained from the old ones through linear 
relations with constant coefficients. That is
\begin{equation}
\bar a^{\sss M} = L(a^{\sss M}),\mbox{ and } \bar j^{\sss M} = L(j^{\sss M}),
\end{equation}
where $L$ is a 6x6 matrix with constant entries. These are detailed in appendix \ref{AppRelAa-O}. As a consequence the fields $\bar a^{\sss M}$ and $\bar j^{\sss M}$ satisfy (\ref{BoxN-aN}) as well and 
the propagation formula (\ref{propagation-a}) is the same. Finally, the field $A^{\sss H}$ follows the same propagation formula, namely (\ref{source-initial-A-intro}) as announced before.


\appendix

\section{Explicit formulas}\label{AppExplicit}

\subsection{Minkowskian systems of coordinates}\label{AppSystem}

These systems $\{x_{\sss M}^\mu\}$  are obtained 
from the system (\ref{coordMinkGene-xHxN}) whith $x^+_{\sss M}=1$ and $x_{\sss M}^c=0$. 
The Minkowskian charts cover the regions $f_{\sss M} > 0$ (Fig. \ref{Fig3}). 
The Weyl factor which relates their corresponding Minkowski spaces ($P_{\sss M} \cap \mathcal{C}$) and 
the de~Sitter space reads respectively
\begin{align*}
K_{\sss N} (x_{\sss N}) &= (1 - H^2\, x^2_{\sss N})^{-1},\\
K_{\sss S} (x_{\sss S}) &= (1 - H^2\, x^2_{\sss S})^{-1},\\
K_{\sss DN} (x_{\sss DN}) &= (H x_{\sss DN})^{-1},\\
K_{\sss DS} (x_{\sss DS}) &= (H x_{\sss DS})^{-1}.
\end{align*}

The relations between the systems North and South are
\begin{equation}\label{transition-ns}
\left \{
 \begin{aligned}
x_{\sss S}^\mu& =  -  \frac{x_{\sss N}^\mu}{H^2 x_{\sss N}^2}\\
x_{\sss N}^\mu& =  -  \frac{x_{\sss S}^\mu}{H^2 x_{\sss S}^2}.
\end{aligned}
\right.
\end{equation}
The relations between the systems Down North and Down South are identical with the 
replacement of the subscripts $N \rightarrow DN$ and $S \rightarrow DS$.

The relations between the systems North and Down North are
\begin{equation}\label{transition_DNN}
\left \{
\begin{aligned}
x_{\sss DN}^0
&= \frac{1/H}{K_{\sss N}(1-Hx^0_{\sss N})-1}  
\\
r_{\sss DN}
&=  \frac{K_{\sss N}r_{\sss N}}{K_{\sss N}(1-Hx^0_{\sss N})-1}.
\end{aligned}
\right.
\end{equation}
The relations between the systems South and Down South are identical with the 
replacement of the subscripts $N \rightarrow S$ and $DN \rightarrow DS$.

We finally mention that the systems North and South  are sometimes called 
polyspherical systems. The Down North and Down South systems are 
conformally flat systems of de~Sitter space. In the notations of \cite{HiguchiCheong-1},
we have $r_{\sss DN} = r$, $x_{\sss DN}^0 = \lambda$ and $r_{\sss DS} = \hat{r}$, 
$x_{\sss DS}^0 = -\hat{\lambda}$.

\subsection{Relations between the fields $a$ and $A$ in the North chart}\label{AppRelAa-N}

The linear relations between the $a^{\sss H}$ and $A^{\sss H}$ in the North chart reads 

\begin{equation}\label{a(A)-H}
\left \{
 \begin{array}{lcl}
{\displaystyle a^{\sss H}_5} &=& {\displaystyle \frac{1}{2K_{\sss N}}
\biggl\{A_c^{\sss H} \left(1 - \frac{x_{\sss N}^2}{4} \right) 
- A_\sigma^{\sss H}x_{\sss N}^\sigma  }\biggr.\\ 
&+&{\displaystyle A_+^{\sss H} K_{\sss N} \left(1 + H^2\right) \biggr\}}\\
{\displaystyle a^{\sss H}_4} &=& {\displaystyle \frac{-1}{2K_{\sss N}}
\biggl\{A_c^{\sss H} \left(1  + \frac{x_{\sss N}^2}{4} \right) }
+{\displaystyle   A_\sigma^{\sss H}x_{\sss N}^\sigma  }\biggr.\\ 
&-&{\displaystyle A_+^{\sss H} K_{\sss N} \left(1 - H^2\right) \biggr\}} 
\\
 {\displaystyle a^{\sss H}_\mu} &=& {\displaystyle \frac{1}{2K_{\sss N}}
\left\{A_c^{\sss H}  \eta_{\mu\nu}x_{\sss N}^{\mu} + 2 A_\mu^{\sss H} \right\}}  .
 \end{array}
\right.
\end{equation}

These relations are obtained by first expressing the basis $\{dy\}$ in the left hand side of (\ref{decomposition-a-AH}) in terms of the basis $\{dx_{\sss N}^c, dx_{\sss N}^\mu, dx_{\sss H}^+\}$ 
and identifying both sides and then applying the correspondence (\ref{tildeGeneral}).

The relations between the $a^{\sss N}$ and the $A^{\sss N}$ are obtained in the same way from the
decomposition (\ref{decomposition-a-AM}) in the basis 
$\{dx_{\sss N}^c, dx_{\sss N}^\mu, dx_{\sss N}^+\}$, they reads
\begin{equation}\label{a(A)-N}
\left \{
 \begin{array}{lcl}
{\displaystyle a^{\sss N}_5} &=& {\displaystyle +\frac{1}{2}
\biggl\{A_c^{\sss N} \left(1 - \frac{x_{\sss N}^2}{4} \right) }
-{\displaystyle   A_\sigma^{\sss N} x_{\sss N}^\sigma +  A_+^{\sss N} \biggr\}}
\\
{\displaystyle a^{\sss N}_4} &=& {\displaystyle -\frac{1}{2}
\biggl\{A_c^{\sss N} \left(1  + \frac{x_{\sss N}^2}{4} \right) }
+{\displaystyle   A_\sigma^{\sss N}x_{\sss N}^\sigma -  A_+^{\sss N} \biggr\}} 
\\
 {\displaystyle a^{\sss N}_\mu} &=& {\displaystyle \frac{1}{2}
\left\{A_c^{\sss N}  \eta_{\mu\nu}x_{\sss N}^{\mu} + 2 A_\mu^{\sss N} \right\}}  .
 \end{array}
\right.
\end{equation} 
The same relations hold for the two sets of current fields $j^{\sss H}$, $J^{\sss H}$ 
and $j^{\sss N}$, $J^{\sss N}$.

Remark that in the North chart for which the Minkowskian plane $P_{\sss N}$ is nothing but 
the plane $P_{\sss H}$ for $H=0$, the above relations can also be obtained  from (\ref{a(A)-H})
setting $H=0$ which implies $K_{\sss N} = 1$ and replacing the superscript $H$ by the superscript $N$.

As already shown in \cite{pconf3} the relation (\ref{tildeGeneral}) implies the Weyl relations
$a^{\sss H} =\left(K_{\sss N}\right)^{-1} a^{\sss N}$
between the two realizations of the same field $\tilde a$ on de Sitter and Minkowski spaces.
As a consequence, the relations (\ref{a(A)-H}) and (\ref{a(A)-N}) leads to, what we call, the extended Weyl transformation between 
$A^{\sss H}$ and $A^{\sss N}$:
\begin{equation}\label{ExtendedWeylA}
\begin{cases}
&{\displaystyle A^{\sss H}_c = A_c^{\sss N} - H^2K^{\sss N} A_+^{\sss N}}\\
&{\displaystyle A^{\sss H}_\mu = A^{\sss N}_\mu + \frac{1}{2} \eta_{\mu\nu} x_{\sss N}^\nu} H^2K^{\sss N} A^{\sss N}_+\\ 
&{\displaystyle A^{\sss H}_+ = A^{\sss N}_+}.
\end{cases}
\end{equation}
Setting the constraint $A^{\sss N}_+ = A^{\sss H}_+ = 0$ leads to the complete equivalence between desitterian fields $A^{\sss H}$ and their Minkowskian counterpart $A^{\sss N}$. 

Now, from (\ref{a(A)-N}) one has 
\begin{equation}\label{AmuAc-N}
\left \{
\begin{array}{lcl}
A^{\sss N}_c &=& a^{\sss N}_5 -  a^{\sss N}_4
\\
A^{\sss N}_{\mu} & = & a_{\mu}^{\sss N} - \frac{1}{2} \eta_{\mu\nu} x_{\sss N}^\nu(a^{\sss N}_5 -  a^{\sss N}_4)
\end{array}
\right .
\end{equation} 
which, under the constraint $A^{\sss N}_+ = A^{\sss H}_+ = 0$, applies also to 
desitterian fields $A^{\sss H}$. 
Thus,
\begin{equation}
\left \{
 \begin{array}{lcl}
  A_c &=& a^{\sss N}_5 -  a^{\sss N}_4
\\
A_{\mu} & = & a_{\mu}^{\sss N} - \frac{1}{2} \eta_{\mu\nu} x_{\sss N}^\nu(a^{\sss N}_5 -  a^{\sss N}_4)
\\
  A_{+} & = & 0 
\end{array}
\right .
\end{equation} 
where we have omitted the superscripts $H$ or $N$ on the fields $A$ since they are equal when 
the constraints apply.

For future comparisons we rewrite these formulas 
using the coordinates $u=y^5+y^4$ and $v=y^5-y^4$ which are 
convenient in calculations
\begin{equation}\label{A(a)N}
\left \{
 \begin{array}{lcl}
  A_c &=& 2a^{\sss N}_v
\\
A_{\mu} & = & a_{\mu}^{\sss N} -  \eta_{\mu\nu} x_{\sss N}^\nu a^{\sss N}_v
\\
  A_{+} & = & 0.
\end{array}
\right .
\end{equation} 
The same considerations apply to the currents as well.

\subsection{Relations between the fields $a$ and $A$ in the other Minkowkian charts}\label{AppRelAa-O}
In each chart South, Down North and Down South, relations analogous to (\ref{A(a)N}) can be obtained. From these, one could derive, using the procedure of Sec. \ref{SecPropFor-N} an expression of the propagation formula in each chart. 
Instead, as explained in Sec. \ref{SecPropGeneral}, an easier way is to remark 
that the fields $a^{\sss M}$ can be redefined through linear combinations with constant coefficients in such a way that the same expression (\ref{A(a)N}), and thus the same propagation formula, apply. 
\subsubsection*{Relations in the South chart}
The relation corresponding to (\ref{A(a)N}) in the South chart reads
\begin{equation}\label{A(a)S}
\left \{
 \begin{array}{lcl}
  A_c &=& 2H^2a^{\sss S}_u
\\
A_{\mu} & = & a_{\mu}^{\sss S} - H^2 \eta_{\mu\nu} x_{\sss S}^\nu a^{\sss S}_u
\\
  A_{+} & = & 0.
\end{array}
\right .
\end{equation} 
From it we can find the transformation $\bar a=L(a)$ which reads
\begin{equation}
\bar a^{\sss S}_u= H^{-2} a^{\sss S}_v, \mbox{ and } \bar a^{\sss S}_v= H^2 a^{\sss S}_u.
\end{equation}

\subsubsection*{Relations in the Down North chart}
In the Down North chart the relation (\ref{A(a)N}) is replaced by
\begin{equation}\label{A(a)DN}
\left \{
 \begin{array}{lcl}
  A_c &=& -2(H^2a^{\sss DN}_u+a^{\sss DN}_v+Ha^{\sss DN}_0)
\\
A_{0} & = &{\displaystyle \frac{1}{H}(H^2a^{\sss DN}_u-a^{\sss DN}_v)}\\
      &-&{\displaystyle x^0_{\sss DN}(H^2a^{\sss DN}_u+a^{\sss DN}_v+Ha^{\sss DN}_0)}
\\
A_{i} & = &a^{\sss DN}_i- \delta_{ij}x^j_{\sss DN}(H^2a^{\sss DN}_u+a^{\sss DN}_v+Ha^{\sss DN}_0)
\\
  A_{+} & = & 0.
\end{array}
\right .
\end{equation} 
From inspection, one deduces that the linear combinations
\begin{equation}
\left \{
 \begin{array}{lcl}
\bar a^{\sss DN}_v&=& (H^2a^{\sss DN}_u+a^{\sss DN}_v+Ha^{\sss DN}_0)\\
  \bar a^{\sss DN}_0&=& {\displaystyle \frac{1}{H}(H^2a^{\sss DN}_u-a^{\sss DN}_v) }\\
  \bar a^{\sss DN}_i&=& a^{\sss DN}_i,
 \end{array}
\right .  
\end{equation}
gives again a system of the form (\ref{A(a)N}).

\subsubsection*{Relations in the Down South chart}
In this chart the relations 
\begin{equation}\label{A(a)DS}
\left \{
 \begin{array}{lcl}
  A_c &=& -2(a^{\sss DS}_u-H^2a^{\sss DS}_v+Ha^{\sss DS}_0)
\\
A_{0} & = &{\displaystyle\frac{1}{H}(a^{\sss DS}_u+H^2a^{\sss DS}_v)}\\
&+&{\displaystyle x^0_{\sss DS}(a^{\sss DS}_u-H^2a^{\sss DS}_v+Ha^{\sss DS}_0)}
\\
A_{i} & = &a^{\sss DS}_i+ \delta_{ij} x^j_{\sss DS}(a^{\sss DS}_u-H^2a^{\sss DS}_v+Ha^{\sss DS}_0)
\\
  A_{+} & = & 0,
\end{array}
\right .
\end{equation} 
replaces (\ref{A(a)N}). The form of that system is recovered for
\begin{equation}
\left \{
 \begin{array}{lcl}
\bar a^{\sss DS}_v&=& -(a^{\sss DS}_u-H^2a^{\sss DS}_v+Ha^{\sss DS}_0)\\
  \bar a^{\sss DS}_0&=& {\displaystyle \frac{1}{H}(a^{\sss DS}_u+H^2a^{\sss DS}_v)} \\
  \bar a^{\sss DS}_i&=& a^{\sss DS}_i.
 \end{array}
\right .  
\end{equation}

\section{Propagation formula}\label{AppPropGen}
In this section we prove a result similar to the one obtained in \cite{HiguchiCheong-1} in a somewhat more abstract manner.

Let $M$ be a globally hyperbolic space time and $A_I$ a family of fields, the dynamics of which, in absence of charge, depends on a Lagrangian density ${\cal L}(A_I,\partial_\mu A_I)$. Suppose moreover that this Lagrangian is a real homogeneous polynomial of degree 2. Let us consider a source represented by the current $J^I$. Taking into account the source, the Lagrangian reads
\begin{equation}
{\cal L}_s={\cal L}+A_IJ^I.
\end{equation}

We claim that the knowledge of $J$ and of the values of $A_I$ and $\partial_\mu A_I$ on a Cauchy hyper-surface allows to compute $A_I$ anywhere on $M$. See \cite{HiguchiCheong-1} for another proof of a similar result.

We define
\begin{equation}\label{def}
\left(L_\pi A\right)^{\mu I}:=\frac{\partial {\cal L}}
{\partial\left(\partial_\mu A_I\right)}
\mbox{ and }
M (A)^{ I}:=\frac{\partial {\cal L}}
{\partial\left( A_I\right)}.
\end{equation}
The lagrangian $\cal L$ being a quadratic form, one can define the unique symmetric bilinear form $(\ ,\ )$ such that
\begin{equation}\label{symform}
{\cal L}(A)=(A,A).
\end{equation}
Namely, this bilinear form is defined through
\begin{equation}
(A,B):=\frac{1}{2}({\cal L}(A+B)-{\cal L}(A)-{\cal L}(B)).
\end{equation}
From (\ref{symform}) one can write
\begin{equation}
{\cal L}(A+h)-{\cal L}(A)= 2(A,h)+ o(h),
\end{equation}
from which, one obtains, using (\ref{def}),
the following useful result:
\begin{equation}\label{bilsym}
\left(L_\pi A\right)^{\mu I}\partial_\mu B_I+M(A)^IB_I=2(A,B).
\end{equation}
Let us consider the current
\begin{equation}\label{divergence}
{\cal J}^\mu(A,B)=A_I\left(L_\pi B\right)^{\mu I}-B_I\left(L_\pi A\right)^{\mu I}.
\end{equation}
Let us first remark that a straightforward calculation using (\ref{bilsym}) gives
\begin{equation}
\begin{split}
\nabla_{\mu}{\cal  J}^\mu(A,B)=
A_I\left[\nabla_\mu\left(L_\pi B\right)^{\mu I}-M(B)^I\right]\\
-B_I
\left[\nabla_\mu\left(L_\pi A\right)^{\mu I}-M(A)^I\right],
\end{split}
\end{equation}
in which one can recognize the Euler-Lagrange equation associated to the Lagrangian $\cal L$. 
 As a result, use of Stokes formula proves that the new symmetric form
\begin{equation}
\langle A, B \rangle=-i\int_\Sigma \sigma_\mu {\cal J}^\mu(A,B^*),
\end{equation}
is an hermitian  sesquilinear form which does not depend on the Cauchy space-like hyper-surface $\Sigma$ as soon as $A$ and $B$ are solutions of the Euler-Lagrange equation. 

Assume now that $A_I$ is solution of the equation with source term:
\begin{equation}
\partial_\mu\left(L_\pi A\right)^{\mu I}=M(A)^I+J^I,
\end{equation}
and $G_{II'}^{R}(x,x')$ is the retarded Green function, that is the two-points function satisfying
\begin{equation}
\partial_\mu\left(L_\pi G_{I}^{R }(x,\ )\right)^{\mu I'}(x')=M(G_{I}^{R }(x,\ ))^{I'}(x')+\delta^{I'}_{I}\delta(x,x'),
\end{equation}
and whose support (related to $x$) lies in the past cone of $x'$. Using (\ref{divergence}) and the Stokes formula, one obtains at once
\begin{align}
A_{I}(x)&=\int_{D_+(\Sigma)}G_{II'}^{R}(x,x')J^{I'}(x'){\rm d}v(x')\nonumber\\
&\ \  -\int_{\Sigma}\sigma'_\mu {\cal J}^\mu\left(A,G_{I}(x,\ )\right),\label{GeneralPropa}\\
&:=(G*J)_{I}(x)+\langle A, G_{I}^{R}(x,\ )\rangle\\
&:= A^{(s)}_{I}(x)+A^{(i)}_{I}(x),
\end{align}
where $\Sigma$ is some Cauchy surface, $D_+(\Sigma)$ the future of $\Sigma$ and $x$ some point lying in $D_+(\Sigma)$.


\bigskip

\begin{thebibliography}{AAA} \baselineskip=10pt
\bibitem{pconf3} S. Faci, E. Huguet, J. Queva, and J. Renaud, Phys. Rev. D {\bf 80}, 124005 (2009).
\bibitem{HiguchiCheong-1} A. Higuchi and L. Y. Cheong, Phys. Rev. D {\bf 78}, 084031 (2008).
\bibitem{Woodard} R.P. Woodard, arXiv:gr-qc/040802.
\bibitem{HiguchiCheong-2} A. Higuchi, L. Y. Cheong and J. R. Nicholas, Phys. Rev. D {\bf 80}, 107502 (2009).
\bibitem{AllenJacobson} B. Allen, T. Jacobson, Comm. Math. Phys.,  {\bf 103}, 669 (1986). 
\bibitem{EastwoodSinger} M. Eastwood and M. Singer, Phys. Lett. {\bf 107A}, 73 (1985).
\bibitem{pconf1} E. Huguet, J. Queva, and J. Renaud, Phys. Rev. D {\bf 73}, 084025 (2006).
\bibitem{pconf2} E. Huguet, J. Queva, and J. Renaud, Phys. Rev. D {\bf 77}, 044025 (2008).
\bibitem{BarutED} A.O.~Barut, \emph{Electrodynamics and classical theory of fields and particles}, Dover Publications, inc. New York (2010).
\bibitem{BrillDeser} D. Brill and S. Deser, Commun. Math. Phys. {\bf 32}, 291 (1973).
\bibitem{Penrose} R. Penrose, in {\it Relativity Groups and Topology, Les Houches 1963}, edited by C. de Witt 
and B. de Witt (Gordon and Breach, New York, 1964).
\bibitem{Bicak-Krtous-1} Bi\u{c}\'{a}k and Krtous, Phys. Rev. D {\bf 64}, 12420 (2001).
\bibitem{kastrup} H.A. Kastrup, AnnalenPhys. {\bf17}, 631 (2008).
\bibitem{BirrelDavies} N.D. Birrell and P.C.W. Davies, \emph{Quantum fields in curved space}, (Cambridge University Press, Cambridge, 1982).
\bibitem{Dirac6cone} P.A.M. Dirac, Ann. Math. {\bf 37}, 429 (1936).
\bibitem{MS} G.~Mack and A.~Salam, Ann. Phys. (N.Y.) {\bf 53}, 174 (1969).
\bibitem{Bayenetal} F.~Bayen, M.~Flato, C. Fronsdal, A.~Haidari, Phys. Rev. D {\bf 32}, 2673 (1985). 


\end{thebibliography}
\end{document}